\documentclass[twocolumn,aps,prl,superscriptaddress,amsmath,amssymb,superscriptaddress]{revtex4-2}
\usepackage[english]{babel}
\usepackage[utf8]{inputenc}
\usepackage[colorinlistoftodos, color=green!40, prependcaption]{todonotes}
\usepackage{amsthm}
\usepackage{mathtools}
\usepackage{siunitx} 
\usepackage{mhchem}
\usepackage{braket}
\usepackage{physics}
\usepackage{xcolor}
\usepackage{graphicx}
\usepackage{adjustbox}
\usepackage{placeins}
\usepackage[T1]{fontenc}
\usepackage{lipsum}
\usepackage{csquotes}\usepackage[pdftex, pdftitle={Article}, pdfauthor={Author}]{hyperref} 
\begin{document}
\title{Enhanced Electron Spin Coherence in a GaAs Quantum Emitter}

\author{Giang~N.~Nguyen}
\email{giang.nguyen@unibas.ch}
\affiliation{Department of Physics, University of Basel, 4056 Basel, Switzerland}
\author{Clemens~Spinnler}
\affiliation{Department of Physics, University of Basel, 4056 Basel, Switzerland}
\author{Mark~R.~Hogg}
\affiliation{Department of Physics, University of Basel, 4056 Basel, Switzerland}
\author{Liang~Zhai}
\affiliation{Department of Physics, University of Basel, 4056 Basel, Switzerland}
\author{Alisa~Javadi}
\altaffiliation[Current address: ]{School of Electrical and Computer Engineering, University of Oklahoma, Norman, OK 73019, USA}
\affiliation{Department of Physics, University of Basel, 4056 Basel, Switzerland}
\author{Carolin~A.~Schrader}
\affiliation{Department of Physics, University of Basel, 4056 Basel, Switzerland}
\author{Marcel~Erbe}
\affiliation{Department of Physics, University of Basel, 4056 Basel, Switzerland}
\author{Marcus~Wyss}
\affiliation{Swiss Nanoscience Institute, University of Basel, 4056 Basel, Switzerland}
\author{Julian~Ritzmann}
\affiliation{Lehrstuhl für Angewandte Festkörperphysik, Ruhr-Universität Bochum, 44780 Bochum, Germany}
\author{Hans-Georg~Babin}
\affiliation{Lehrstuhl für Angewandte Festkörperphysik, Ruhr-Universität Bochum, 44780 Bochum, Germany}
\author{Andreas~D.~Wieck}
\affiliation{Lehrstuhl für Angewandte Festkörperphysik, Ruhr-Universität Bochum, 44780 Bochum, Germany}
\author{Arne~Ludwig}
\affiliation{Lehrstuhl für Angewandte Festkörperphysik, Ruhr-Universität Bochum, 44780 Bochum, Germany}
\author{Richard~J.~Warburton}
\affiliation{Department of Physics, University of Basel, 4056 Basel, Switzerland}
\date{\today} 

\begin{abstract}
A spin-photon interface should operate with both coherent photons and a coherent spin to enable cluster-state generation and entanglement distribution.
In high-quality devices, self-assembled GaAs quantum dots are near-perfect emitters of on-demand coherent photons. However, the spin rapidly decoheres via the magnetic noise arising from the host nuclei. Here, we address this drawback by implementing an all-optical nuclear-spin cooling scheme on a GaAs quantum dot. The electron-spin coherence time increases 156-fold from $T_2^* = \SI{3.9}{\nano\s}$ to $\SI{0.608}{\micro\s}$. The cooling scheme depends on a non-collinear term in the hyperfine interaction. The results show that such a term is present even though the strain is low and no external stress is applied. 
Our work highlights the potential of optically-active GaAs quantum dots as fast, highly coherent spin-photon interfaces.
\end{abstract}

\maketitle

\section*{Introduction}
Photonic quantum technologies require quantum emitters capable of high-fidelity and high-rate operation. Of particular interest for quantum networks \cite{wehner_quantum_2018,pompili_realization_2021,stas_robust_2022,azuma_allphotonic_2015,buterakos_deterministic_2017,borregaard_oneway_2020} and measurement-based quantum computing \cite{raussendorf_oneway_2001,browne_resourceefficient_2005,gimeno-segovia_threephoton_2015a} are quantum emitters that host a spin \cite{atature_material_2018a}, allowing the creation of spin-photon entanglement.

Self-assembled semiconductor quantum dots (QDs) are potential candidates for spin-photon interfaces due to the deterministic photon emission at exceptionally high quality and rates \cite{somaschi_nearoptimal_2016a,wang_optimal_2019,uppu_scalable_2020,tomm_bright_2021a} and the ability to load a QD with a single electron or hole \cite{warburton_single_2013a}. This has led to demonstrations of spin-photon entanglement \cite{degreve_quantumdot_2012,gao_observation_2012,schaibley_demonstration_2013,appel_entangling_2022}, distant spin-spin entanglement \cite{delteil_generation_2016a,stockill_phasetuned_2017c}, and the creation of multi-photon cluster states \cite{schwartz_deterministic_2016,cogan_deterministic_2023a,coste_highrate_2023}. 
However, in these previous experiments, the short spin coherence times, $T_2^*\sim1-\SI{10}{\nano\s}$, limited the entanglement fidelity. The short $T_2^*$ is a consequence of magnetic noise in the host nuclear spins, coupling to the electron spin via the hyperfine interaction \cite{merkulov_electron_2002, cywinski_pure_2009,coish_hyperfine_2004}. 


A powerful way to mitigate the short $T_2^*$ is to cool the nuclear spins to ultralow temperatures in order to reduce the fluctuations. The nuclei can be cooled via the electron spin itself, exploiting the hyperfine interaction \cite{yang_general_2013}. In an optical experiment, this was originally demonstrated on an ensemble of QDs \cite{greilich_nucleiinduced_2007}. On single QDs, nuclear spin cooling was demonstrated on gate-defined GaAs QDs via a measure-and-correct feedback loop \cite{shulman_suppressing_2014b,nakajima_coherence_2020}. More recently, the highly inhomogeneous nuclear spins of a self-assembled InGaAs QD were cooled via an autonomous feedback \cite{gangloff_quantum_2019a}. Subsequently, a quantum sensing protocol was employed, narrowing the nuclear distribution further, thereby increasing $T_2^*$ to \SI{300}{\nano\s}~\cite{jackson_optimal_2022}. For both schemes, a non-collinear term in the hyperfine interaction is required to allow for the cooling of the nuclei. In contrast to the collinear term from the contact hyperfine interaction ($\propto S_z I_z$), the non-collinear term ($\propto S_z I_x$) arises from nuclear quadrupolar fields in strained QDs; here $S_z$ ($I_z$) is the electron (nuclear) spin operator along the direction of the applied magnetic field 
\cite{huang_theoretical_2010,yang_general_2013,hogele_dynamic_2012}.




The most studied QDs for spin-photon applications are QDs in the InGaAs/GaAs system. InGaAs QDs are self-assembled via the strain-driven Stanski-Krastanov mechanism. Self-assembled GaAs QDs in an AlGaAs matrix represent an alternative platform. The strain is low such that these QDs are self-assembled via an alternative mechanism, droplet-etching. Low-noise GaAs QDs have excellent photonic properties, all at a convenient wavelength (around 780~nm). In high-quality material, the optical linewidths are within 10\% of the transform limit \cite{zhai_lownoise_2020}. Photons emitted by remote QDs have achieved a two-photon interference visibility of 93\% without spectral or temporal filtering \cite{zhai_quantum_2022}. The biexciton cascade generates entangled photon pairs with an extremely high entanglement concurrence \cite{reindl_allphotonic_2018}. In terms of the nuclear spins, the lack of both strain and spin-$\frac{9}{2}$ In atoms results in a homogeneous nuclear spin ensemble \cite{chekhovich_nuclear_2020}, as demonstrated by the success of the Carr-Purcell-Meiboom-Gill (CPMG) decoupling scheme in prolonging the electron spin $T_2$ from \SI{3.8}{\micro\s} to \SI{113}{\micro\s} \cite{zaporski_ideal_2023a}. However, as for InGaAs QDs, noise in the nuclear spins limits $T_2^*$ to values of a few-ns. To date, the possibility of feedback cooling the nuclear spins via the electron spin has remained uncertain, due to the predicted absence of the strain-generated non-collinear hyperfine interaction.

\begin{figure*}[ht]  \centering
  \includegraphics[width=1\textwidth]{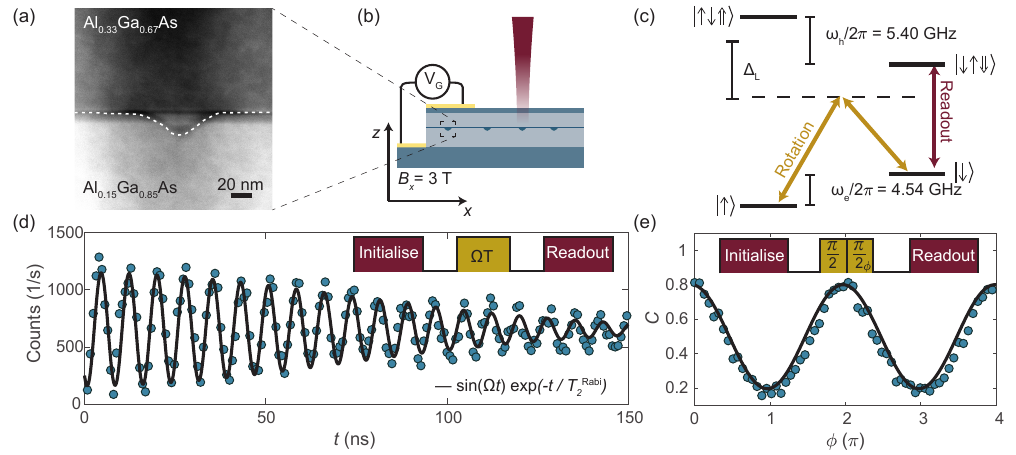}
  \caption{Coherent spin control of an electron in a droplet-etched GaAs QD. (a) High-angle dark-field scanning transmission image of a droplet-etched GaAs QD. The dashed line is a guide to the eye to describe the droplet shape. (b)~Schematic of the sample design: a layer of GaAs QDs is embedded in a diode structure. A magnetic field perpendicular to the growth direction defines the quantization axis. (c) Energy level diagram of a charged QD in an in-plane magnetic field. The ``vertical'' transitions are $x$-polarised while the ``diagonal'' transitions are $y$-polarised. A circularly polarised rotation pulse detuned by $\Delta_{\text{L}} = \SI{700}{\giga\Hz}$ drives a Raman transition between the electron spin states. The readout laser is on resonance with the lower-frequency ``vertical'' transition and initialises the electron into the $\ket{\uparrow}$-state. (d) Electron spin Rabi oscillations as a function of drive time $t$. The solid line is an exponential fit to the data with $T_2^{\text{Rabi}} = \SI{73\pm5}{\nano\s}$. (e) Full control of the rotation axis about the Bloch sphere using two consecutive $\frac{\pi}{2}$-pulses as a function of the phase $\phi$ of the second pulse. The solid line is a sinusoidal fit to the data.}
  \label{fig:1}
\end{figure*}

Here, we implement all-optical cooling schemes on low-noise GaAs QDs and demonstrate an increase in the electron spin coherence time from $T_2^* = \SI{3.9}{\nano\s}$ to $\SI{0.608}{\micro\s}$. 
This is achieved with autonomous feedback and without any external perturbation (such as strain tuning).
We demonstrate spin control with \mbox{$T_2^* = \SI{0.608}{\micro\s}$}, an extension of $T_2$ with CPMG (with a scaling of $T_2^{\text{CPMG}} \propto N^{0.69}$ matching previous experiments \cite{zaporski_ideal_2023a}), fast spin rotations (Rabi frequencies above 100~MHz), and high-fidelity spin control (\mbox{$F_\pi>98\%$}). Our results establish GaAs QDs as an emitter of coherent photons and a host to a coherent spin. 

To create the QDs, droplet-etched nanoholes in an $\text{Al}_{0.15}\text{Ga}_{0.85}\text{As}$ matrix are filled with GaAs and capped by an $\text{Al}_{0.33}\text{Ga}_{0.67}\text{As}$ layer. The materials are almost lattice-matched. Figure~\ref{fig:1}(a) shows a high-angle dark-field scanning transmission (HAADF-STEM) image of a GaAs QD \cite{_sup_}. Notable is a thin, Al-rich layer at the bottom surface of the QD \cite{_sup_}. The QD is embedded in a p-i-n diode structure (see Fig.~\ref{fig:1}(b)) such that the QD charge is stabilised via the Coulomb blockade. Individual QDs exhibit near-transform-limited optical linewidths \cite{_sup_,zhai_lownoise_2020,zhai_quantum_2022}. A $\SI{3.00}{\tesla}$ magnetic field is applied perpendicular to the growth direction (Voigt geometry), at an angle of $45^\circ$ to the in-plane crystal axes. The electron Zeeman frequency is \mbox{$f_{\text{Z}}=\SI{4.54}{\giga\Hz}$} corresponding to a g-factor of \mbox{$g_\text{e}=-0.11$}.

The spin is manipulated by a two-colour Raman pulse detuned from the excited states by \mbox{$\Delta_{\text{L}} = \SI{700}{\giga\Hz}$} (see Fig. \ref{fig:1}(c)). This pulse is created by amplitude-modulating circularly-polarised light with an electro-optic modulator driven by an arbitrary waveform generator \cite{_sup_,bodey_optical_2019a}.
A laser resonant with the red ``vertical'' transition is used to read out the spin (such that the $\ket{\downarrow}$-state is bright, the $\ket{\uparrow}$-state is dark) and to prepare the spin in the $\ket{\uparrow}$-state via optical spin pumping \cite{_sup_}.

Driving the electron spin resonance (ESR) (Fig.~\ref{fig:1}(d)) shows clear Rabi oscillations between $\ket{\uparrow}$ and $\ket{\downarrow}$ with increasing drive time $t$. 
We find an exponential decay of the oscillations with \mbox{$T_2^{\text{Rabi}} = \SI{73\pm5}{\nano\s}$}, corresponding to a quality factor of \mbox{$Q = 2T_2^{\text{Rabi}}f_{\text{Rabi}} = \SI{19\pm1}{}$} and $\pi$-pulse fidelity \mbox{$f_\pi = \frac{1}{2}(1+e^{-1/Q}) = \SI{0.975\pm0.002}{}$} at \mbox{$\Omega=2\pi\times\SI{130}{\mega\Hz}$}. 
As has been observed for InGaAs QDs \cite{bodey_optical_2019a}, we find a strong modulation of the quality factor \cite{_sup_} when the electron spin is driven close to the nuclear Larmor frequencies $\omega_\text{n}$ (i.e., $\Omega\sim\omega_\text{n}$), a signature of an electron-nuclei interaction via a Hartmann-Hahn resonance \cite{hartmann_nuclear_1962}.

We access rotation around a second axis on the Bloch sphere by controlling the phase of the microwave signal that is imprinted on the optical field. Fig.~\ref{fig:1}(e) shows the sinusoidal response after two consecutive $\frac{\pi}{2}$-pulses on changing the phase $\phi$ of the second pulse, thereby demonstrating rotation around an arbitrary axis on the equator of the Bloch sphere. 

\begin{figure*}[ht]
  \centering
  \includegraphics[width=1\textwidth]{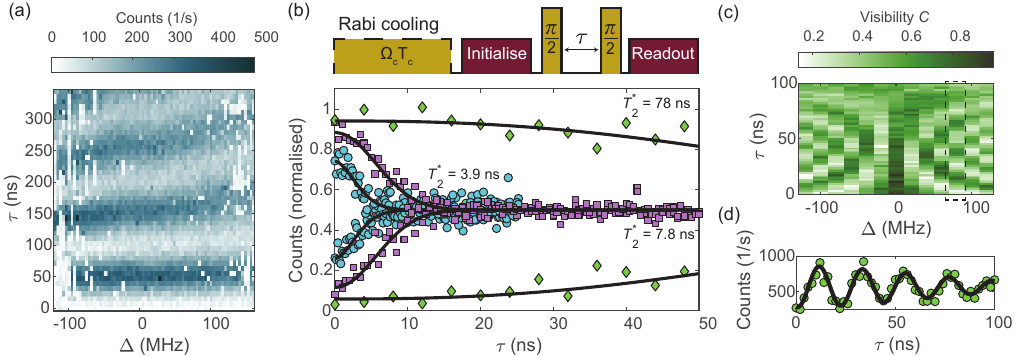}
  \caption{Locking of electron spin resonance (ESR) and cooling of nuclei with a Rabi drive. (a) Rabi oscillations versus detuning show locking of the ESR to the drive within a window of frequencies and unstable Rabi oscillations outside the window. (b) Top: Pulse sequence for Ramsey interferometry with prior Rabi cooling. For Rabi cooling a $T_\text{c} = \SI{1}{\micro\s}$ long pulse at a Rabi frequency of $\Omega_\text{c} = 2\pi\times \SI{17}{\mega\Hz}$ is used. The Ramsey experiment was performed at a larger Rabi frequency of $2\pi\times \SI{100}{\mega\Hz}$. Bottom: Top and bottom envelopes of the Ramsey interferometry with $\SI{100}{\micro\s}$ pause (circles), zero pause (squares) and Rabi cooling (diamonds); the extracted coherence times are \mbox{$T_2^* = \SI{3.9\pm0.2}{\nano\s}$}, \mbox{$T_2^* = \SI{7.8\pm0.2}{\nano\s}$}, and \mbox{$T_2^* = \SI{78\pm2}{\nano\s}$}, respectively. Counts are normalised to $0.5$ for long delays. (c) Ramsey interferometry as a function of detuning with respect to the cooling frequency $f_\text{c}$ of the Rabi drive. (d) Linecut at \mbox{$\Delta = \SI{50}{\mega\Hz}$} with \mbox{$T_2^* = \SI{87\pm6}{\nano\s}$} (dashed box in (c)). The solid lines in (b) and (d) are Gaussian fits to the data.}
  \label{fig:2}
\end{figure*}

On driving Rabi oscillations as a function of the detuning $\Delta$ with respect to the Zeeman frequency (\mbox{$\Delta = f_{\text{Z}}-f_{\text{probe}}$}), we find strong deviations from the typical chevron pattern expected for a two-level system (see Fig.~\ref{fig:2}(a)). In a $\sim \SI{200}{\mega\Hz}$ window around the Zeeman frequency, we find that the spin rotations lock to the probe frequency $f_{\text{probe}}$, a clear signature of electron spin--nuclear spin coupling \cite{xu_optically_2009a,kloeffel_controlling_2011,hogele_dynamic_2012,urbaszek_nuclear_2013a}.

When the ESR is locked via the hyperfine interaction, cooling of the nuclei, equivalently narrowing of the nuclear distribution, is predicted \cite{vink_locking_2009a,latta_confluence_2009a}. This can be quantified by a reduction in $\sigma_{\text{OH}}$, the standard deviation of the ESR frequency fluctuations due to the changing Overhauser field. To probe this, we perform a free-induction decay (FID) experiment to measure the electron coherence time $T_2^*$ in a Ramsey experiment, which acts as a gauge of the temperature of the nuclear spin ensemble ($\sigma_\text{OH}\propto T_2^*$) \cite{cywinski_pure_2009, coish_hyperfine_2004}. 
We compare the bare $T_2^*$ to that obtained after locking the ESR (see Fig.~\ref{fig:2}(b)). We observe a 20-fold increase from \mbox{$T_{2}^* = \SI{3.9\pm0.2}{\nano\s}$} to $\SI{78\pm2}{\nano\s}$ corresponding to a narrowing of $\sigma_{\text{OH}}$ from $\SI{52\pm1}{\mega\Hz}$ to $\SI{2.90\pm0.05}{\mega\Hz}$ following the Rabi drive. 
Remarkably, we already find an enhancement in coherence time without a dedicated cooling pulse when the Ramsey experiment is carried out with a high duty cycle: repetitive Ramsey experiments lead to a $T_2^*$ of $\SI{7.8\pm0.2}{\nano\s}$. To determine the bare electron coherence time, we add a $\SI{100}{\micro\s}$ buffer between each cycle. This observation suggests that the repetitive application of spin manipulation pulses as short as $\SI{4}{\nano\s}$ already leads to a narrowing of $\sigma_{\text{OH}}$. 

We confirm the nuclear-spin cooling and locking of the ESR to the Rabi drive by fixing the cooling frequency $f_{\text{c}}$ during Rabi cooling, subsequently detuning the probe frequency $f_{\text{probe}}$ in a Ramsey experiment. Oscillations arise at the detuning frequencies $\Delta = f_{\text{c}}-f_{\text{probe}}$ as expected in a classic Ramsey experiment (see Fig.~\ref{fig:2} (c,d)), now with an increased coherence time.

\begin{figure*}[ht]  \centering
  \includegraphics[width=1\textwidth]{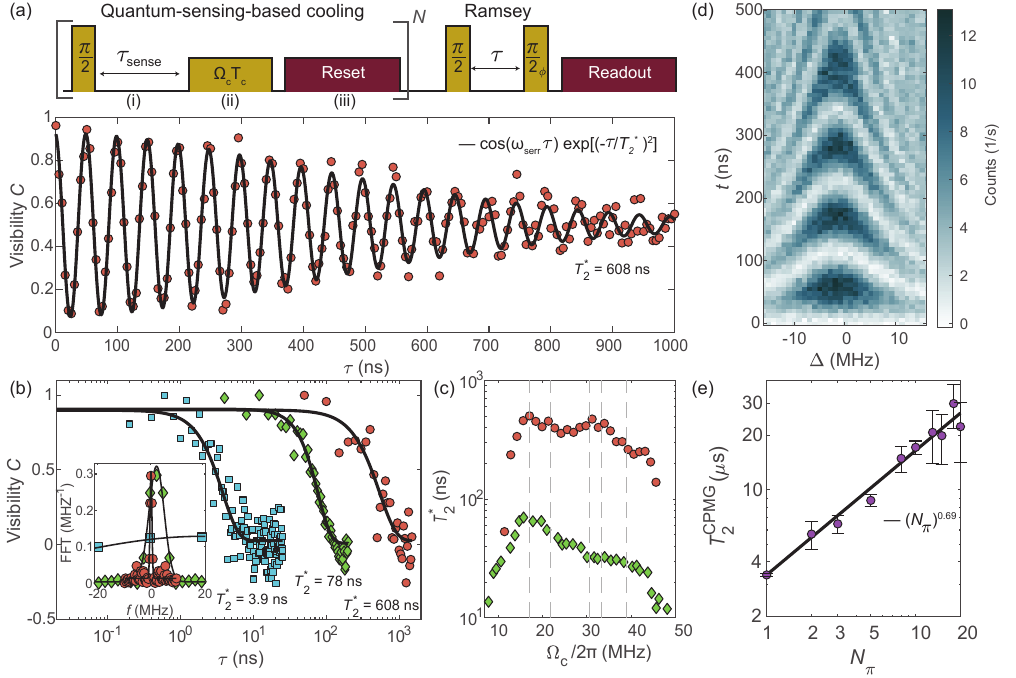}
  \caption{Quantum-sensing-based cooling and dynamical decoupling. (a) Top: Pulse scheme for the quantum-sensing-based cooling consisting of (i) a sensing step, (ii) a driven electron-nuclei interaction, and (iii) a reset. The last reset pulse in the cooling scheme initialises the electron spin for the Ramsey experiment performed at a Rabi frequency of $2\pi\times \SI{100}{\mega\Hz}$. Bottom: Ramsey interferometry with serrodyne frequency \mbox{$\omega_{\text{serr}} = 2\pi\times\SI{20}{\mega\Hz}$} ($\phi(\tau) = \sin(\omega_{\text{serr}}\tau)$) following quantum-sensing-based cooling gives $T_2^* = \SI{0.608\pm0.013}{\micro\s}$. 
  (b) Comparison of $T_2^*$ before cooling (squares), after Rabi cooling (diamonds), and after quantum-sensing-based cooling (circles). Inset: fast Fourier transform of the Ramsey visibilities gives \mbox{$\sigma_{\text{\text{OH}}} = \SI{52\pm 1}{\mega\Hz}$}, \mbox{$\sigma_{\text{\text{OH}}} = \SI{2.90\pm0.05}{\mega\Hz}$}, and \mbox{$\sigma_{\text{\text{OH}}} = \SI{0.355\pm0.004}{\mega\Hz}$}, respectively. 
  (c) $T_2^*$ versus Rabi frequency during cooling ($\Omega_{\text{c}}$) for Rabi cooling (diamonds) and quantum-sensing-based cooling (circles). Dashed lines correspond to nuclear Larmor frequencies, from left to right:
  \mbox{$\Delta\omega=2\pi\times\SI{17.08}{\mega\Hz}$}, \mbox{$\omega(\ce{^{75}As})=2\pi\times\SI{21.9}
  {\mega\Hz}$}, \mbox{$\omega(\ce{^{69}Ga})=2\pi\times\SI{30.7}
  {\mega\Hz}$}, \mbox{$\omega(\ce{^{27}Al})=2\pi\times\SI{33.28}
  {\mega\Hz}$}, \mbox{$\omega(\ce{^{71}Ga})=2\pi\times\SI{39.0}{\mega\Hz}$}. 
  (d) Rabi oscillations at \mbox{$\Omega_{\text{c}} = 2\pi\times\SI{8.9}{\mega\Hz}$} as a function of detuning from $f_\text{c}$ following quantum-sensing-based cooling. (e) Dynamical decoupling of the electron spin with a CPMG sequence. The solid lines in (a,b) are Gaussian fits to the data. The solid line in (e) is a power law fit to the data.}
  \label{fig:3}
\end{figure*}

To cool the nuclei further, we implement the recently developed quantum-sensing-based cooling scheme \cite{jackson_optimal_2022}. In this protocol, each cooling cycle consists of three steps (see Fig.~\ref{fig:3}(a, top)): (i) The electron spin is initialised and then rotated to the equator with a $\frac{\pi}{2}$-pulse. A period of free evolution $\tau_\text{sense}$ allows the electron to sense the Overhauser field fluctuation that leads to a detuning $\Delta$ from the target frequency $f_\text{c}$.
(ii) A coherent electron-nuclei flip-flop interaction arising from a non-collinear term in the hyperfine interaction is activated through ESR driving at Hartmann-Hahn resonance $\Omega\approx \omega_\text{n}$. The sign of the detuning $\Delta$ determines the direction of the nuclear flops and thus leads to a reversal of the measured fluctuation. (iii) A projective measurement of the spin state transfers entropy from the nuclei and concludes one cycle of the cooling scheme. Repeating this cycle with increasing sensing time $\tau$ results in a narrower feedback function in each cycle and hence an increased sensitivity to changes in $\sigma_{\text{OH}}$.

We find optimal parameters for the quantum-sensing-based cooling at \mbox{$N=40$} cycles with a linearly increasing sensing time $\tau_\text{sense}$ from $\tau_{\text{min}} = \SI{20}{\nano\s}$ to \mbox{$\tau_{\text{max}} = \SI{400}{\nano\s}$}, and electron-nuclei drive time \mbox{$T_{\text{c}} = \SI{125}{\nano\s}$} at a Rabi frequency \mbox{$\Omega_{\text{c}} = 2\pi\times\SI{17}{\mega\Hz}$}, followed by a spin pumping pulse of $\SI{200}{\nano\s}$ \cite{_sup_}. This preparation sequence takes $\sim\SI{22}{\micro\s}$ and is repeated before each Ramsey cycle. 

\newpage

The electron coherence time $T_2^*$ increases from $\SI{3.9\pm0.2}{ns}$ to $\SI{0.608\pm0.013}{\micro\s}$ after application of the protocol (see Fig.~\ref{fig:3}(a, b)). This constitutes a 156-fold increase in $T_2^*$. The final $T_2^*$ is a factor of two larger than the previous highest $T_2^*$ reported on an electron spin hosted by an InGaAs QD (\SI{296}{\nano\s} \cite{jackson_optimal_2022}) and just below the highest reported $T_2^*$ of a single electron spin qubit in a gate-defined GaAs QD (\SI{767}{\nano\s} \cite{nakajima_coherence_2020}). The enhancement corresponds to a narrowing of the nuclear-spin ensemble from \mbox{$\sigma_{\text{OH}} = \SI{52\pm1}{\mega\Hz}$} to \SI{0.355\pm0.004}{\mega\Hz} (see Fig.~\ref{fig:3}(b, inset)).

Using hyperfine constants $A_k$ and abundancies $\eta_k$ of the nuclei species $k\in\{\ce{^{69}Ga},\ce{^{71}Ga},\ce{^{75}As}\}$ we can estimate the number of nuclei involved \mbox{$N = 5/4\sum_k\eta_kA_k^2T_2^{*2} = 1.4\cdot 10^5$} and estimate the hyperfine interaction per nuclei \mbox{$A_\text{c} = 1/(\sqrt{5N/2}\pi T_2^{*}) = \SI{0.13}{\mega\Hz}$} \cite{merkulov_electron_2002,jackson_optimal_2022,zaporski_ideal_2023a}. This corresponds to a distribution of $\sigma_{\text{OH}}/A_\text{c} \approx 376.8$ macrostates in the uncooled state and 2.6 after quantum-sensing-based cooling, entering the regime where just a few nuclei excitations remain.

For both the quantum-sensing-based and Rabi cooling schemes, the Rabi frequency $\Omega_{\text{{c}}}$ is an important parameter (see Fig. \ref{fig:3}(c)). The maximum performance for both cooling schemes occurs at \mbox{$\Omega_{\text{{c}}} =2\pi\times \SI{17}{\mega\Hz}$}, close to the difference frequency of $\ce{^{71}Ga}$ and $\ce{^{75}As}$ (\mbox{$\Delta\omega=\omega(\ce{^{71}Ga})-\omega(\ce{^{75}As}) = 2\pi\times\SI{17.08}{\mega\Hz}$}). This result is in contrast to those on InGaAs QDs for which cooling was most effective at a direct Hartmann-Hahn resonance \cite{jackson_optimal_2022}. Generally speaking, the fact that cooling via an autonomous feedback process is effective on GaAs QDs shows that a non-collinear term in the hyperfine interaction \cite{hogele_dynamic_2012,xu_optically_2009a,yang_general_2013} must be present even though the strain in the QDs is small.

Following cooling, a typical chevron pattern is observed on driving Rabi oscillations as a function of detuning with respect to the cooling frequency $f_{\text{c}}$ (Fig.~\ref{fig:3}(d)), using here a Rabi frequency below the Hartmann-Hahn resonances. This demonstrates that in this case the electron spin is isolated from the nuclear environment and behaves as a two-level system. 
In addition, the quality factor of the oscillations now increases to \mbox{$Q=\SI{30\pm1.4}{}$} (corresponding to a $\pi$-pulse fidelity of $\SI{98.4\pm0.1}{\percent}$) \cite{_sup_}, consistent with a reduction of hyperfine-interaction-induced Rabi decay. 

Recent experiments showed that the electron spin $T_2$ can be increased by implementing a decoupling scheme, the CPMG protocol. As a final step, we verify that this is also possible on the QD for which nuclear spin cooling was highly effective (see Fig. \ref{fig:3}(d)). By applying CPMG pulses, we extend $T_2$ from \mbox{$T_2^{\text{HE}} = \SI{2.93\pm0.06}{\micro\s}$} using a Hahn echo ($N_\pi$=1) to \mbox{$T_2^{\text{CPMG}} = \SI{22\pm8}{\micro\s}$}, an order of magnitude increase, with \mbox{$N_\pi$=20} pulses. We extract a $T_2$ scaling of $T_2^{\text{CPMG}}\propto N_\pi^\gamma$ with \mbox{$\gamma = \SI{0.69\pm0.12}{}$}, consistent with recent results on droplet-etched QDs \cite{zaporski_ideal_2023a} and gate-defined QDs \cite{malinowski_spectrum_2017}. This result confirms that the nuclear spin ensemble is highly homogeneous. The application of more pulses is currently limited by imperfect pulse calibrations and the electron spin relaxation time \mbox{$T_1\sim\SI{40}{\micro\s}$} \cite{_sup_}.

In conclusion, we have demonstrated fast and flexible optical control of an electron spin confined to a self-assembled GaAs QD. We show that autonomous feedback protocols to cool the nuclear spins are very effective even on an as-grown, close-to-strain-free QD. Nuclear-spin cooling leads to a 156-fold increase in the $T_2^*$ time, \mbox{$T_2^* = \SI{0.608}{\micro\s}$}. Furthermore, both $T_2^*$ and $T_2$ can be extended on exactly the same QD, $T_2^*$ by nuclear spin cooling, $T_2$ by dynamic decoupling. These results imply that a small non-collinear term must be present in the hyperfine Hamiltonian. Following nuclear spin cooling, $T_2^*$ becomes much longer than both the time required to rotate the spin and the time required to generate a photon. Together with recent results on the generation of indistinguishable photons from remote GaAs QDs \cite{zhai_quantum_2022} performed on the same sample as used in this experiment, our results highlight the promise of GaAs QDs for a coherent spin-photon interface. Furthermore, the system represents an ideal testbed for creating non-classical collective states within the nuclear spin ensemble \cite{zaporski_manybody_2023a}.


We thank Ming-Lai Chan and Peter Lodahl at the Niels-Bohr Institute, Leon Zaporski and Mete Atat\"{u}re at the University of Cambridge, and Dorian Gangloff at the University of Oxford for stimulating discussions.

The work was supported by SNF Project 200020\_204069 and Horizon 2020 FET-Open Project QLUSTER. LZ, GNN and AJ received funding from the European Union’s Horizon 2020 Research and Innovation Programme under the Marie Sk\l{}odowska-Curie grant agreement No.\ 721394 (4PHOTON), No.\ 861097 (QUDOT-TECH), and No.\ 840453 (HiFig), respectively. HGB, JR, ADW and AL acknowledge financial support from the grants DFH/UFA CDFA05-06, DFG TRR160, DFG project 383065199, and BMBF Q.Link.X 16KIS0867.
\bibliographystyle{apsrev4-2} 
\bibliography{GaAsSpin.bib} 

\newpage
\onecolumngrid

\section*{Supplementary Information: Enhanced Electron Spin Coherence in a GaAs Quantum Emitter}
\setcounter{section}{0}
\section{Experimental Setup}

\begin{figure*}[ht]    \centering
    \includegraphics[width=1\textwidth]{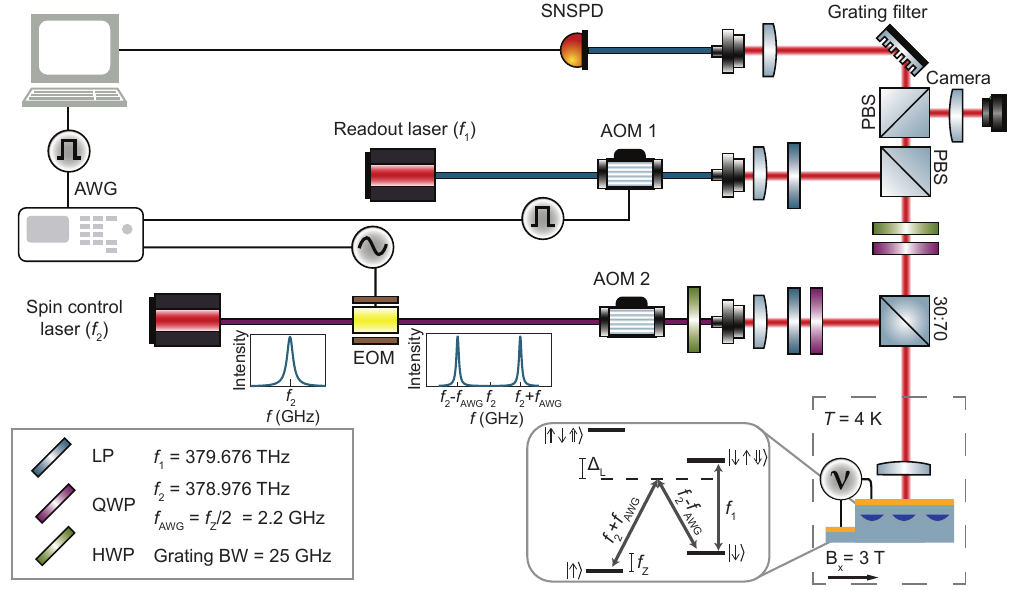}
    \caption{\textbf{Experimental setup.} Spin manipulation is achieved using optical pulses at frequency $f_2$, detuned from $f_1$ by $\Delta_\text{L}$. Optical sidebands at $f_2\pm f_\text{AWG}$ are generated by an EOM that is modulated at half the Zeeman frequency (\mbox{$f_{\text{AWG}}=f_\text{Z}/2$}). A readout laser at frequency $f_1$ is used for spin initialisation (via optical spin pumping) and spin readout. Back-reflected laser light at $f_1$ is suppressed via a cross-polarisation microscope head consisting of two polarising beam splitters (PBS), a linear polariser (LP), and a quarter-wave plate (QWP). Laser light at $f_2$ is rejected with a grating filter with $\SI{25}{\giga\Hz}$ bandwidth (BW) centred around $f_1$. A half-wave plate (HWP) before the QD is used to match the polarisation of the excitation to the ``vertical'' QD transition. }
    \label{suppfig:setup}
\end{figure*}

The quantum dot (QD) sample is cooled down to $\SI{4.2}{\K}$ in a helium bath cryostat (see Fig.~\ref{suppfig:setup}). A \mbox{$B=\SI{3.00}{\tesla}$} magnetic field is applied perpendicular to the QD growth direction (Voigt geometry), at $45^\circ$ to the crystal axes (along the [100] or [010] direction). QD excitation, spin manipulation, and readout are performed all-optically with the use of a dark-field microscope \cite{kuhlmann_darkfield_2013}. QD emission is filtered from the excitation laser at frequency $f_1$ (Toptica DL Pro) using two polarising beam-splitters (PBS). Spin initialisation and readout are triggered using an acousto-optic modulator (AOM, Gooch\&Housego 3200-124). Spin manipulation is facilitated with a second laser at frequency $f_2$, \mbox{$\Delta_\text{L}=\SI{700}{\giga\Hz}$} red detuned from $f_1$. The laser is amplitude-modulated with an electro-optic modulator (EOM, Jenoptik AM785) that is driven at $f_{\text{AWG}}$ to generate sidebands at $\pm f_{\text{AWG}}$. A second AOM (Gooch\&Housego 3200-1113) is used to set the amplitude of the spin control pulses. A quarter-wave plate (QWP) sets the required circular polarisation and the pulse is sent down the axis of the cryostat to the QD with a $30:70$ beam splitter. The reflected rotation laser light is filtered from the QD emission with a $\SI{25}{\giga\Hz}$-bandwidth grating filter.
Control signals for spin initialisation, manipulation and readout are generated by an arbitrary waveform generator (AWG, Tektronix 7122C) and QD counts are detected with a superconducting nanowire single photon detector (SNSPD, Single Quantum) and processed with a time-tagger (Swabian Instruments Time Tagger Ultra). The QD sample is the same as used in Ref.~\cite{zhai_lownoise_2020} and \cite{zhai_quantum_2022} and growth details can be found in Ref.~\cite{babin_charge_2021}.

\section{Quantum dot}

\subsection{Structural properties and composition}
We perform energy-dispersive X-ray spectroscopy (EDX) on a QD from the same wafer to determine the spatial distribution of arsenic, gallium and aluminium atoms. The sample preparation for EDX/scanning transmission electron microscopy (STEM) was carried out in an FEI Helios NanoLab 650 DualBeam, a combined scanning electron microscope (SEM) and focused ion beam (FIB). A double layer of carbon is deposited to protect the QD from ion-induced damage. The first C-layer was deposited using electron-induced deposition at a beam energy of $\SI{5}{\kilo\eV}$ and a beam current of $\SI{3.2}{\nano\A}$. The second C-layer was deposited with ion-induced deposition at a beam energy of $\SI{30}{\kilo\eV}$ and a beam current of $\SI{83}{\pico\A}$. Sample cutting and polishing were carried out with the FIB at a beam energy of $\SI{30}{\kilo\V}$ and beam currents ranging from $\SI{240}{\pico\A}$ down to $\SI{83}{\pico\A}$. The sample thickness in the upper area, where the QD is located, was $<\SI{50}{\nano\m}$. The imaging of the TEM specimen was carried out in a JEOL JEM-F200 operated in the STEM-mode at a beam energy of \SI{200}{\kilo\V}. A high-angle dark-field scanning transmission image (HAADF-STEM) of a QD is shown in Fig.~1 of the main text.

\begin{figure*}[ht]    \centering
    \includegraphics[width=1\textwidth]{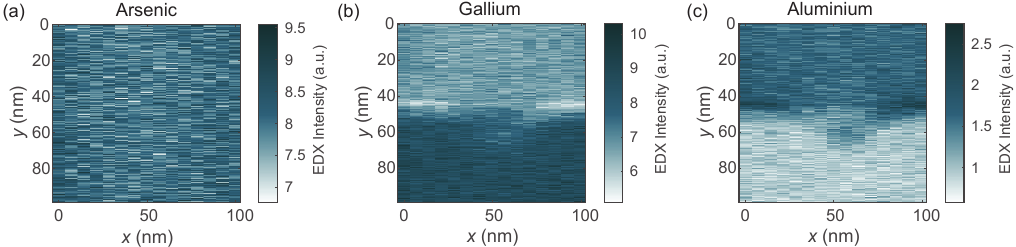}
    \caption{\textbf{Energy-dispersive x-ray spectroscopy.} EDX intensity as a function of x-and y- position around the QD for (a) arsenic, (b) gallium, and (c) aluminium atoms. }
    \label{suppfig:EDX}
\end{figure*}

The arsenic EDX intensity is homogeneous as expected: the arsenic concentration is constant throughout the growth of the QD layer and matrix material (see Fig.~\ref{suppfig:EDX}(a)). The EDX intensity for gallium atoms shows a high signal below the QD ($y\sim\SI{50}{\nano\m},x\sim\SI{50}{\nano\m}$) and a low signal above (see Fig.~\ref{suppfig:EDX}(b)). This is expected as the matrix material below the QD is Al$_{0.15}$Ga$_{0.85}$As and the QD is grown on and capped with Al$_{0.33}$Ga$_{0.67}$As. Accordingly, the QD alone would be expected to have an even higher EDX intensity, as it is filled purely with GaAs. However, in the EDX experiment the gallium signal at the QD is dominated by matrix material around the QD. The EDX signal for aluminium atoms shows a low aluminium signal below the QD and a high signal above (see Fig.~\ref{suppfig:EDX}(c)). In addition, a thin film of high aluminium signal can be seen above the QD at $y\sim\SI{45}{nm}$ revealing the presence of a high-aluminum content layer. This layer is formed from aluminium used in drilling the nanoholes. Interestingly, this thin, high Al-content layer also forms at the boundary of the nanohole, leading to the growth of GaAs not on Al$_{0.33}$Ga$_{0.67}$As but on a thin layer with higher Al-content. This high Al-content layer represents an increase in both alloying and strain with respect to a pure GaAs QD \cite{zaporski_ideal_2023a}.

\clearpage
\subsection{Optical properties at zero magnetic field}
\begin{figure*}[ht]    \centering
    \includegraphics[width=1\textwidth]{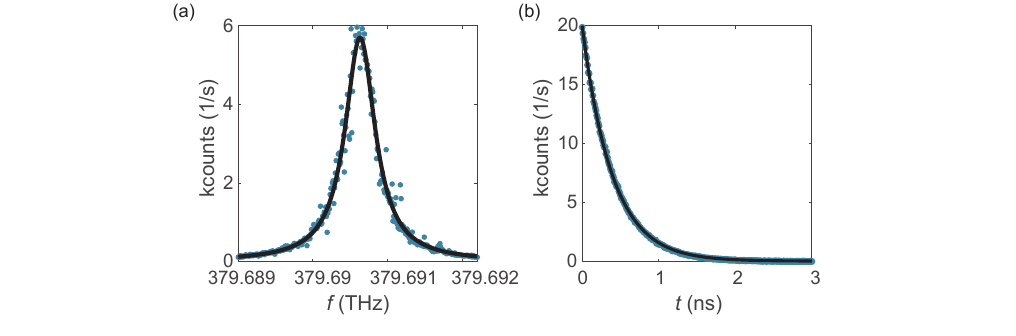}
    \caption{\textbf{Optical properties at zero magnetic field.} (a) Scan of a narrow-band laser through the resonance of the negatively charged exciton $X^-$. The solid line is a Lorentzian fit with FWHM of $\SI{491}{\mega\Hz}$. (b) Lifetime measurement of the $X^-$. The solid line is an exponential fit with lifetime \mbox{$\tau=\SI{392}{\pico\s}$}. This corresponds to a Fourier-limited linewidth of \SI{405}{\mega\Hz}.}
    \label{suppfig:optical prop}
\end{figure*}

The QD linewidth is measured by slowly scanning a narrow-band laser through the $X^-$ resonance (see Fig.~\ref{suppfig:optical prop}(a)). Low excitation power is used to avoid power broadening. A Lorentzian fit to the data gives a FWHM of $\SI{491}{\mega\Hz}$. This should be compared to the Fourier-limited linewidth as inferred from an $X^-$ lifetime measurement. In a pulsed experiment, the emission decay time is recorded in a histogram (see Fig.~\ref{suppfig:optical prop}(b)). An exponential fit gives a $X^-$ lifetime of \mbox{$\tau=\SI{392}{\pico\s}$}, which corresponds to a Fourier-limited linewidth of \mbox{$\frac{1}{2\pi\tau}=\SI{405}{\mega\Hz}$}. The linewidth is therefore $\SI{22}{\percent}$ above the Fourier limit. This is slightly higher than the average QD-linewidth on the sample \cite{zhai_lownoise_2020}. 

\subsection{Optical properties at \textit{B}\,=\,3.00\,T}

\begin{figure*}[ht]    \centering
    \includegraphics[width=1\textwidth]{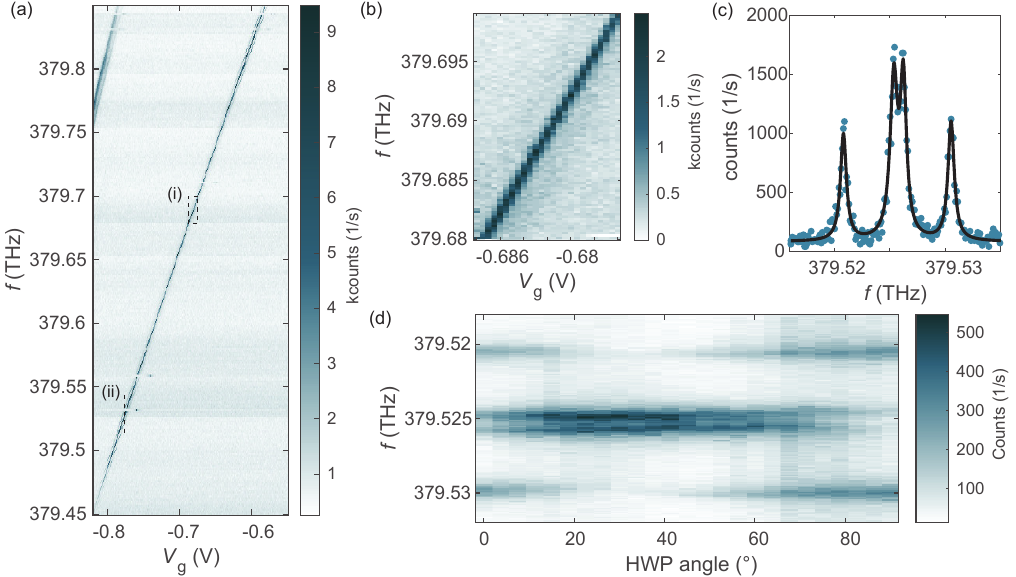}
    \caption{\textbf{Optical properties at \mbox{\textit{B}\,=\,3.00\,T}.} (a) RF plateau map of the $X^-$. (b) Zoom in on the plateau centre, area (i) in (a). The two outer ``vertical'' transitions disappear due to spin pumping, while spin pumping for the inner ``diagonal'' transitions is ineffective due to a too-small frequency splitting. (c) Linescan at (ii) of (a) in the cotunneling regime shows four transitions. Spin-pumping is ineffective due to cotunneling of the electron to the Fermi sea. The solid line is a fit to a sum of four Lorentzians. (d) Linecut at (ii) as a function of excitation polarisation given by the rotation angle of the HWP. }
    \label{suppfig:optical prop 3T}
\end{figure*}

We record a resonance fluorescence (RF) plateau map by scanning both excitation laser frequency ($f$) and gate voltage ($V_\text{g}$) across the $X^-$ transitions at \mbox{$B= \SI{3.00}{\tesla}$} (see Fig.~\ref{suppfig:optical prop 3T}(a)). A fine scan across the plateau centre (see Fig.~\ref{suppfig:optical prop 3T}(b)) shows a single line, while a linecut across the plateau edge (see Fig.~\ref{suppfig:optical prop 3T}(c)) shows four distinct peaks corresponding to the four optical transitions. From a fit to a sum of four Lorentzians, we can extract the ground-state and excited-state Zeeman splittings to be \mbox{$\omega_\mathrm{e}=2\pi\times \SI{4.54}{\giga\Hz}$} and\mbox{ $\omega_\mathrm{h}=2\pi\times\SI{5.40}{\giga\Hz}$}, respectively, corresponding to g-factors (\mbox{$g_\mathrm{e/h}=\hbar\omega_\mathrm{e/h}/\mu_\mathrm{B}B$}) of \mbox{$g_\mathrm{e}=-0.11$} and \mbox{$g_\mathrm{h}=0.13$}. (The assumption here is that $g_\mathrm{e}$ is negative.) Vanishing signal is expected in the plateau centre due to optical spin pumping. The single line we observe arises due to repumping as the laser drives both of the near-degenerate ``diagonal'' transitions. The ``vertical'' transitions are extinguished by spin pumping, as expected. Optical spin pumping is ineffective at the plateau edges due to cotunneling with the Fermi sea. Changing the rotation angle of the HWP in the beam path of the QD excitation (cf.\ Fig.~\ref{suppfig:setup}), we can either address the two outer ``vertical'' transitions (x-polarised), the two inner ``diagonal'' transitions (y-polarised), or all four transitions with ``diagonal'' polarisation (see Fig.~\ref{suppfig:optical prop 3T}(d)). For the spin-manipulation experiments, we set the HWP such that only the outer ``diagonal'' transitions are excited ($\sim 85^\circ$).

\newpage
\section{Spin properties}

\subsection{Data acquisition}
For all experiments, a laser background was recorded by carrying out a measurement with the $X^-$-transition out of resonance with the readout laser. This was carried out by changing the voltage applied to the diode. After subtracting the background signal, counts during the readout pulse were integrated with an integration time depending on the duty cycle of the experiment (typically \SI{5}{}$-$\SI{60}{\second} for a signal to noise ratio of $\sim 4$). For Ramsey and Carr-Purcell-Meiboom-Gill (CPMG) experiments, two measurements were performed. In the second one, the electron is initialised in one state but subsequently projected into the opposite spin-state by adding a $\pi$-phase shift to the final $\frac{\pi}{2}$-pulse. This gives the top and bottom envelopes of the experiment. Additionally, it avoids a dynamic nuclear polarisation from building up \cite{ladd_pulsed_2010,stockill_quantum_2016}. The counts from the two envelopes ($c_{\downarrow}, c_{\uparrow}$) are used to calculate the visibility $C$ via:
\begin{align}
    C=\frac{c_{\downarrow}-c_{\uparrow}}{\text{max}(c_{\downarrow}-c_{\uparrow})}
\end{align}

\subsection{Spin relaxation and optical spin pumping}
\begin{figure*}[ht]    \centering
    \includegraphics[width=1\textwidth]{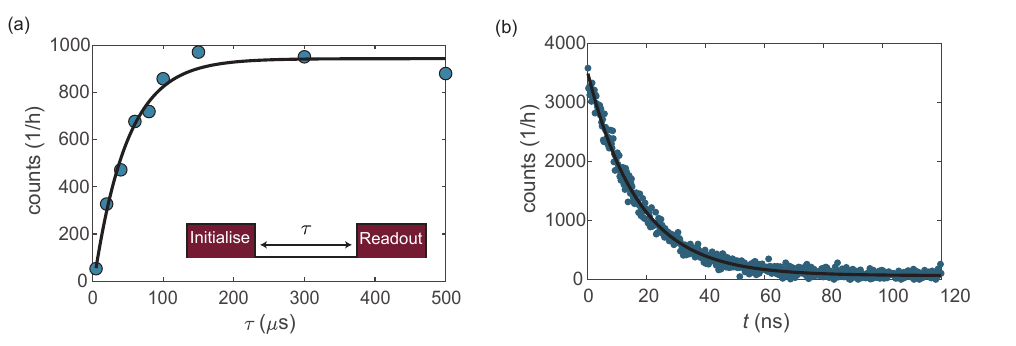}
    \caption{\textbf{Electron-spin relaxation time.} (a) A pump-probe experiment determines the electron relaxation time, \mbox{$T_1=\SI{47\pm7}{\micro\s}$}. (b) Characteristic spin pumping decay. The solid line is an exponential fit with spin pumping time \mbox{$\tau_\text{OSP}=\SI{17\pm0.1}{\nano\s}$} and spin-pumping fidelity \mbox{$F_\text{OSP}=0.99$}.}
    \label{suppfig:T1 OSP}
\end{figure*}

The electron spin lifetime $T_1$ is measured in a pump-probe experiment (see Fig.~\ref{suppfig:T1 OSP}(a)). A $\SI{200}{\nano\s}$ pulse in resonance with the red ``vertical'' transition pumps the spin into the $\ket{\uparrow}$-state. The same pulse is applied after a delay $\tau$. For short delays, no signal is expected from the second readout pulse. For longer delays spin relaxation processes flip the spin back to the $\ket{\downarrow}$-state leading to the reappearance of a signal. By fitting the data to \mbox{$C=C_0(1-\exp(-\tau/T_1))$}, we find an  electron spin lifetime of \mbox{$T_1=\SI{47\pm7}{\micro\s}$}.

Capturing the time histogram of the counts during spin pumping, we find an exponential decay characterising the spin-pumping time (see Fig. \ref{suppfig:T1 OSP}(b)). The spin-pumping time decreases with increasing laser power. For all experiments in this work, the spin pumping laser-power (equivalently, readout laser-power) is set such that the spin pumping time is $\sim\SI{17}{\nano\s}$. An estimate of the spin-pumping fidelity is given by the residual counts ($c_\infty$) compared to the initial counts ($c_0$) via \mbox{$F_\text{OSP}=\sqrt{(1-c_\infty/c_0)}=0.99$} \cite{lobl_narrow_2017a}.\\

\subsection{Rabi frequency scaling}
\begin{figure*}[ht]    \centering
    \includegraphics[width=1\textwidth]{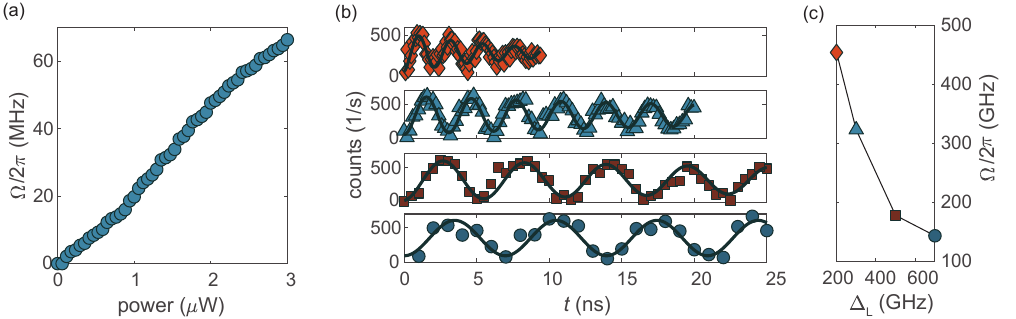}
    \caption{\textbf{Rabi frequency scaling.} (a) Rabi frequency $\Omega/2\pi$ versus optical power follows a linear increase. The laser power is measured after the 30:70 beam splitter in front of the optical window. (b) Rabi oscillations for different laser detunings: from top to bottom \mbox{$\Delta_\text{L}=\SI{200}{\giga\Hz}$}, \mbox{$\Delta_\text{L}=\SI{300}{\giga\Hz}$}, \mbox{$\Delta_\text{L}=\SI{500}{\giga\Hz}$}, \mbox{$\Delta_\text{L}=\SI{700}{\giga\Hz}$}. (c) Rabi frequency as a function of excited state detuning $\Delta_\text{L}$.}
    \label{suppfig:rabi frequency dependencies}
\end{figure*}

Figure \ref{suppfig:rabi frequency dependencies}(a) shows the Rabi frequency $\Omega/2\pi$ of the driven electron spin as a function of laser power. Ideally, $\Omega/2\pi$ depends linearly on the laser power \cite{bodey_optical_2019a}. We find a small deviation to a linear behaviour which we attribute to a slightly nonlinear response of the AOM we used for power control. The Rabi frequency scales inversely with detuning with respect to the excited states $\Delta_\text{L}$. Figure \ref{suppfig:rabi frequency dependencies}(b) shows Rabi oscillations for increasing $\Delta_\text{L}$. The Rabi frequency is extracted from an exponential fit and shown in Figure \ref{suppfig:rabi frequency dependencies}(c). While Rabi frequencies of several hundred\,MHz are possible, the quality factor falls at the highest laser powers on account of a laser-induced spin flip \cite{bodey_optical_2019a}.

\subsection{Laser-induced spin flip}
\begin{figure*}[ht]    \centering
    \includegraphics[width=1\textwidth]{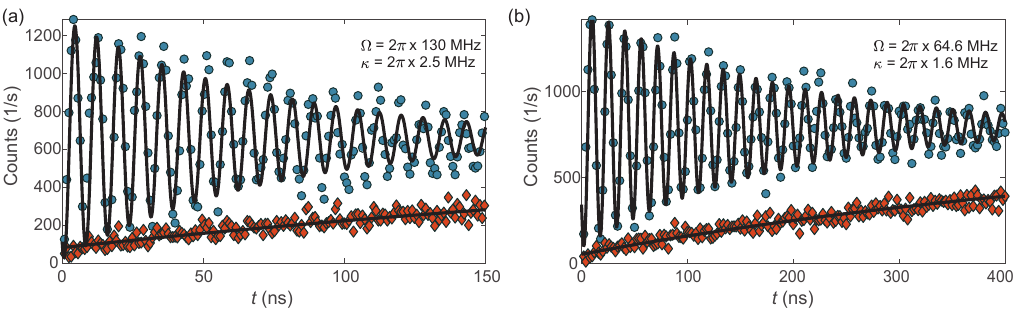}
    \caption{\textbf{Laser-induced spin flip.} Blue circles: Rabi oscillations. Orange diamonds: rotation-laser-induced spin flip on driving Rabi oscillations far off-resonance with respect to the ESR. (a) \mbox{$\Omega=2\pi\times\SI{130}{\mega\Hz}$}, (b) \mbox{$\Omega =2\pi\times \SI{64.6}{\mega\Hz}$}.}
    \label{suppfig:kappa}
\end{figure*}

The effect of a rotation-laser-induced spin flip can be measured by setting $f_{\text{AWG}}$ off-resonance with respect to the electron spin resonance (ESR) (see Fig.~\ref{suppfig:kappa}, diamonds). On increasing the drive time $t$ we see an increase in counts, a consequence of a rotation-laser-induced spin flip. This process limits the quality factor of the Rabi oscillations \cite{bodey_optical_2019a}. An exponential fit $\exp(-\kappa t)$ gives a spin-flip rate of \mbox{$\kappa= 2\pi\times \SI{2.5}{\mega\Hz}$} for \mbox{$\Omega=2\pi\times\SI{130}{\mega\Hz}$} and \mbox{$\kappa= 2\pi\times \SI{1.6}{\mega\Hz}$} for \mbox{$\Omega=2\pi\times\SI{65}{\mega\Hz}$}. We find a scaling factor of \mbox{$\kappa/\Omega=0.19$} and $0.25$, respectively. The mechanism responsible for this unwanted process is unclear. \\

\subsection{Hartmann-Hahn resonances}
\begin{figure*}[ht]    \centering
    \includegraphics[width=1\textwidth]{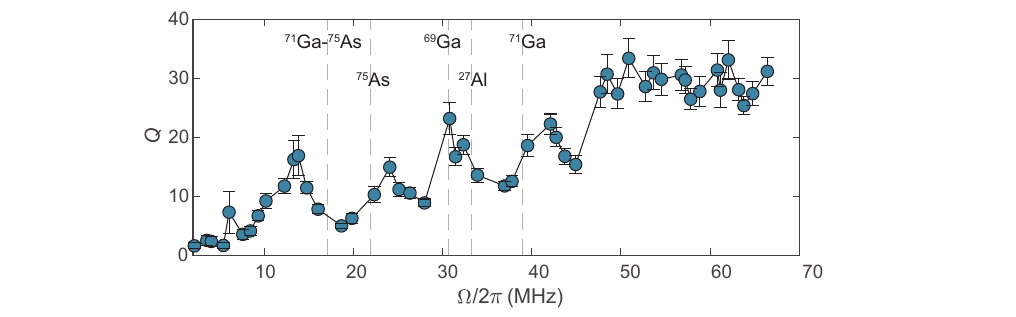}
    \caption{\textbf{Hartmann-Hahn resonances.} Quality factor $Q$ as a function of Rabi frequency $\Omega/2\pi$. For $\Omega>2\pi\times\SI{50}{\mega\Hz}$, $Q$ is constant, $Q\approx30$. Dashed lines correspond to the nuclear Larmor frequencies without quadrupolar shifts.}
    \label{suppfig:hartmann-hahn}
\end{figure*}

The quality factor $Q = 2T_2^{\text{Rabi}}f_{\text{Rabi}}$ of the electron spin rotations shows a nontrivial dependence on the Rabi frequency $f_{\text{Rabi}}=\Omega/2\pi$  (see Fig.~\ref{suppfig:hartmann-hahn}). Driving at $\Omega>2\pi\times\SI{50}{\mega\Hz}$, the quality factor is constant at $Q \approx 30$, limited by rotation-laser-induced spin-flips. For $\Omega<2\pi\times\SI{50}{\mega\Hz}$ there is a complicated $\Omega$-dependence. Hartmann-Hahn resonances are expected when the Rabi frequency matched the nuclei frequencies $\omega_\mathrm{n}$ (i.e., $\Omega\approx\omega_\mathrm{n}$) \cite{hartmann_nuclear_1962}. Something akin to the Hartmann-Hahn process is revealed here as $Q$ rises and falls in the frequency range where $\Omega/2\pi$ and the various values of $\omega_\mathrm{n}/2\pi$ match. However, the minima in $Q$ do not lie consistently at the exact values of $\omega_\mathrm{n}/2\pi$ for the isotopes involved. Furthermore, we also find a minimum in $Q$ close to the difference frequency $^{71}$Ga -- $^{75}$As. Similar structure was observed for InGaAs QDs \cite{bodey_optical_2019a} despite the very different strain environments of the two systems and on gate-defined GaAs QDs \cite{nakajima_coherence_2020}.
\newpage
\section{Cooling of the nuclear ensemble}

\subsection{Rabi cooling}
\begin{figure*}[ht]    \centering
    \includegraphics[width=1\textwidth]{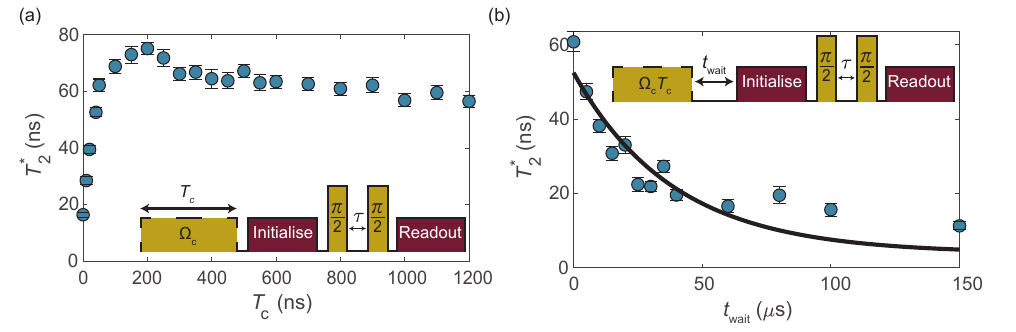}
    \caption{\textbf{Rabi cooling dependencies.} (a) $T_2^*$ as a function of Rabi drive duration $T_{\text{c}}$ before each Ramsey sequence. The entire sequence is repeated many times for each value of $T_\text{c}$. The nuclear spins do not return to their fully warm state ($T_2^*=\SI{3.9}{\nano\s}$) after each sequence. (b) $T_2^*$ decay for increased wait time $t_\text{wait}$ between cooling pulse and Ramsey experiment. The solid line is an exponential fit with a decay time of $\SI{39\pm8}{\micro\s}$ for which we fix the decay to the bare \mbox{$T_2^*=\SI{3.9}{\nano\s}$} for large values of $t_\text{wait}$.}
    \label{suppfig:Rabi cooling}
\end{figure*}

Figure~\ref{suppfig:Rabi cooling}(a) shows the nuclear cooling dependence on the duration of the Rabi cooling pulse length $T_{\text{c}}$. We find a maximum of $T_2^*$ for $T_{\text{c}}\approx \SI{200}{\nano\s}$ and a slight drop of $T_2^*$ for longer $T_{\text{c}}$. Note that a single Rabi cooling cycle is not enough to reach \mbox{$T_2^* = \SI{78}{\nano\s}$} -- it requires the repetitive application of the Rabi cooling pulses and the Ramsey measurement. Adding a waiting time $t_\text{wait}$ between Rabi cooling pulse and the Ramsey experiment, we find a decay of the $T_2^*$ (see Fig.~\ref{suppfig:Rabi cooling}(b)). Fitting the data to a single exponential decay, we find a decay time of $\SI{39\pm8}{\micro\s}$. In both experiments, the initialise/readout pulse is $\SI{200}{\nano\s}$ in duration, the $\frac{\pi}{2}$-pulses are \SI{3}{\nano\s} in duration.

\subsection{Quantum-sensing-based cooling}
\begin{figure*}[ht]    \centering
    \includegraphics[width=1\textwidth]{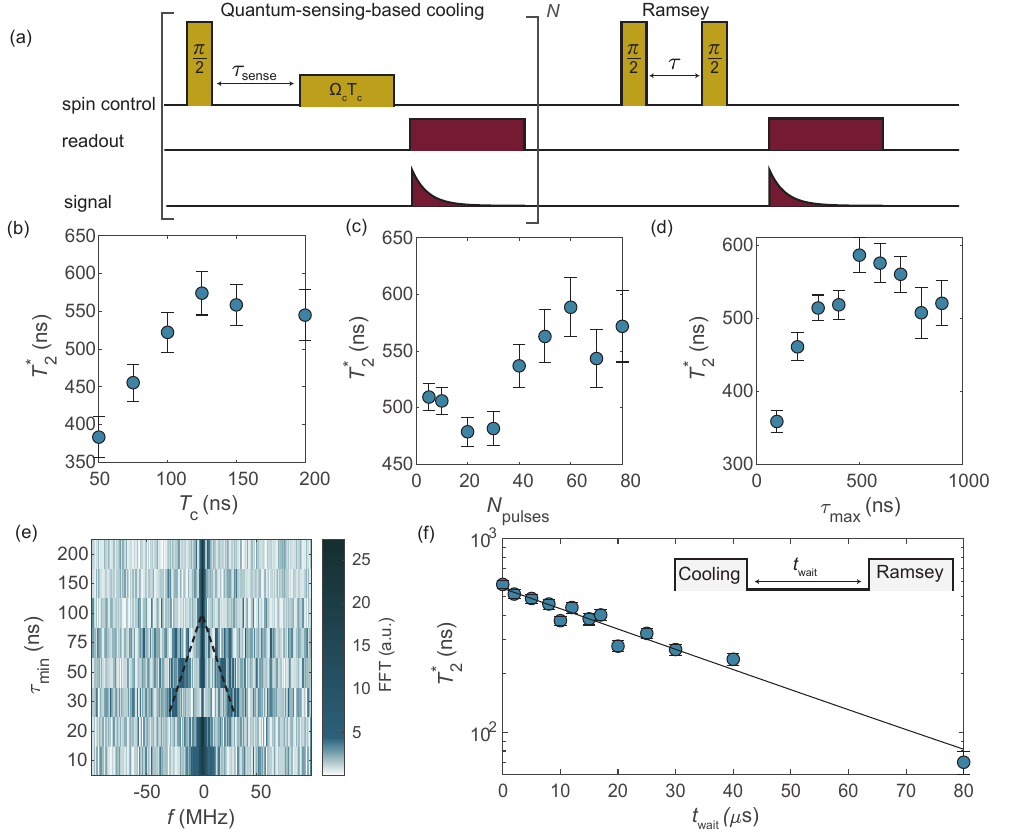}
    \caption{\textbf{Quantum-sensing-based cooling dependencies.} (a) Schematic of pulse sequence for the quantum-sensing-based cooling. The initialise/readout/reset pulse is $\SI{200}{\nano\s}$ in duration, the $\frac{\pi}{2}$-pulses are $\SI{3}{\nano\s}$ in duration. The Rabi frequency during cooling is set to $\Omega_\text{c} = 2\pi\times \SI{17}{\mega\Hz}$. (b) $T_2^*$ as a function of electron-nuclei interaction drive time $T_{\text{c}}$. (c) $T_2^*$ as a function of the number of cooling pulses $N_{\text{pulses}}$. (d) $T_2^*$ as a function of maximum sensing time $\tau_{\text{max}}$. (e) FFT of the Ramsey decay as a function of the minimum sensing time $\tau_{\text{min}}$. Black dashed lines are a guide to the eye to highlight the appearance of side peaks at oscillation frequency $f$ in the Ramsey experiment. (f) $T_2^*$ decay after optimum cooling. The solid line is an exponential fit to the data with a decay time of $\SI{41\pm4}{\micro\s}$ for which we fix the decay to the bare\mbox{ $T_2^*=\SI{3.9}{\nano\s}$} for large values of $t_\text{wait}$.}
    \label{suppfig:feedback cooling}
\end{figure*}

Figure \ref{suppfig:feedback cooling}(a) shows the pulse sequence applied to the spin control laser, the readout laser and the corresponding signal we measure on the SNSPD for the quantum-sensing-based cooling protocol.
Figure \ref{suppfig:feedback cooling}(b-e) summarises the dependencies of the quantum-sensing-based cooling on $T_{\text{c}}$, $N_{\text{pulses}}$, $\tau_{\text{max}}$, and $\tau_{\text{min}}$. Each of the parameters was changed individually with the other parameters kept constant. For $T_{\text{c}}$ and $\tau_{\text{max}}$ a clear maximum could be found at $T_{\text{c}} \approx \SI{125}{\nano\s}$ and $\tau_{\text{max}}\approx\SI{500}{\nano\s}$. Conversely, the dependence on $N_{\text{pulses}}$ did not show a strong optimum, we chose to use $N=40$ pulses. Remarkably, with just \mbox{$N_{\text{pulses}}= 5$} we reach a $T_2^*$ of $\SI{500}{\nano\s}$. This suggests that also the quantum-sensing-based cooling protocol relies on the repetitive application of the cycle in Fig. \ref{suppfig:feedback cooling}(a), i.e., that full cooling is not achieved with a single cycle. $\tau_{\text{min}}$ is set such that the Ramsey envelope does not show oscillations. This is most clearly visible on performing a fast Fourier transform (FFT) on the Ramsey decay. For \mbox{$\tau_{\text{min}}<\SI{30}{\nano\s}$} only a single peak is visible in the FFT. For \mbox{$\tau_{\text{min}}=\SI{30}{\nano\s}$} side peaks appear, suggesting that $\tau_{\text{min}}$ is set too large such that there are multiple locking points \cite{jackson_optimal_2022}. We thus decided to work with \mbox{$\tau_{\text{min}}=\SI{20}{\nano\s}$}.

Figure~\ref{suppfig:feedback cooling}(f) shows $T_2^*$ as a function of the wait time $t_\text{wait}$ between quantum-sensing-based cooling and the Ramsey experiment, similar to Fig.~\ref{suppfig:Rabi cooling}(b). Fitting the data with an exponential decay, we find a decay time of $\SI{41\pm4}{\mu\s}$. The decay time is short compared to the nuclei diffusion times extracted from similar experiments on InGaAs QDs \cite{jackson_optimal_2022} and nuclei diffusion times extracted from NMR experiments on droplet-etched GaAs QDs \cite{chekhovich_nuclear_2020,millington-hotze_nuclear_2023}. Our results on this GaAs QD may hint at the importance of an electron cotunneling or an electron-mediated coupling between separate nuclear spins \cite{wust_role_2016,millington-hotze_nuclear_2023}.

\subsection{Chevron after quantum-sensing-based cooling}
\begin{figure*}[ht]    \centering
    \includegraphics[width=1\textwidth]{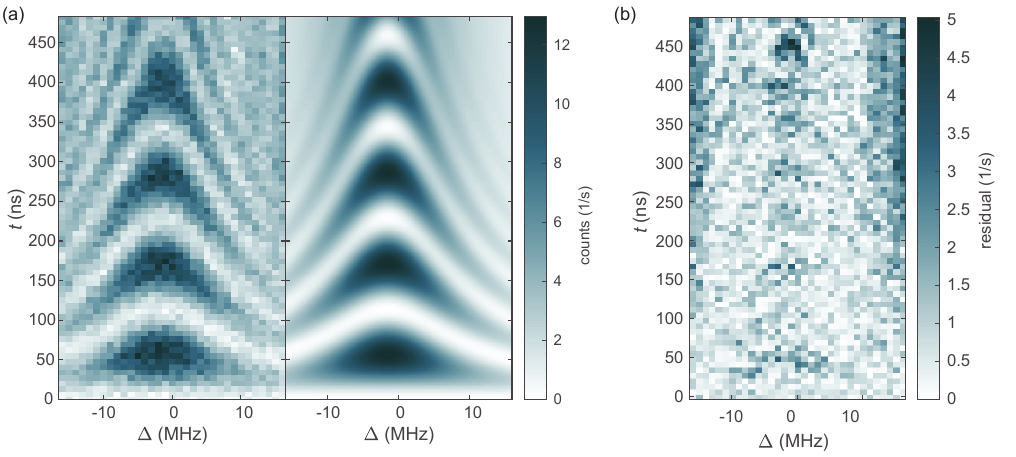}
    \caption{\textbf{Chevron after quantum-sensing-based cooling.} (a) Left: experiment, right: fit of data to Eq.~\ref{Eq: Chevron}. (b) Residual between data and fit.}
    \label{suppfig:Chevron}
\end{figure*}

On measuring Rabi oscillations as a function of ESR detuning (\mbox{$\Delta=f_{\text{AWG}}-f_{\text{Z}}$}) after quantum-sensing-based cooling at \mbox{$f_{\text{c}}=f_{\text{Z}}$} we find the typical chevron pattern (see Fig. \ref{suppfig:Chevron}). Taking into account the effect of Gaussian-distributed Overhauser field noise with effective ESR broadening of width $\sigma_{\text{OH}}$, the Rabi oscillations can be modelled as:
\begin{align}
    C(t,\Delta)=\int_{-\infty}^{+\infty}\text{d}\delta \frac{1}{\sqrt{2\pi}\sigma_{\text{OH}}}\exp{-\frac{\delta^2}{2\sigma_{\text{OH}}^2}}\cdot\frac{\Omega^2}{\Omega^2+(\Delta-\delta_\text{AC}+\delta)^2}\sin^2(\sqrt{\Omega^2+(\Delta-\delta_\text{AC}+\delta)^2)}t/2).
    \label{Eq: Chevron}
\end{align}
Fitting the data to Eq.~\ref{Eq: Chevron} we find a perfect match for \mbox{$\sigma_\text{OH}=\SI{8.1}{\mega\Hz}$} and a small frequency offset of $\delta_\text{AC}=\SI{-1.61}{\mega\Hz}$ (see Fig.~\ref{suppfig:Chevron}). The larger broadening with respect to the fully cooled state can be explained by the fact that the Rabi pulses have a much longer duration than the Ramsey sequence: the narrowed nuclei distribution following quantum-sensing-based cooling is disturbed by the Rabi experiment leading to an increase in $\sigma_\text{OH}$ with respect to the maximally cooled case of \mbox{$\sigma_\text{OH}=\SI{0.36}{\mega\Hz}$} as extracted from a FFT of the Ramsey experiment. 

The frequency offset $\delta_\text{AC}$ arises as a consequence of the AC- Stark effect, which can occur if the couplings of the two transitions of the lambda system are slightly imbalanced (e.g. if the rotation laser polarisation is not perfectly circular). During cooling, the frequency is locked at $f_{\text{c}}+ \delta_{\text{AC}}$, which leads to a small offset in detuning \cite{jackson_quantum_2021}. Here, cooling was performed at $\Omega = 2\pi\times \SI{17}{\mega\Hz}$, while the chevron was recorded at $\Omega = 2\pi\times \SI{8.9}{\mega\Hz}$.

Note that the experimental data were acquired in two runs, the first for Rabi oscillations up to \mbox{$t=\SI{400}{\nano\s}$}, the second for Rabi oscillations from $t=\SI{400}{\nano\s}$ to $\SI{480}{\nano\s}$ in order to resolve the fourth oscillation. This is the origin of the deviation between the model and experiment for \mbox{$t>\SI{400}{\nano\s}$}: there is a slight change in power and cooling performance between the two runs (see Fig.~\ref{suppfig:Chevron}(b)).

\newpage 
\subsection{Rabi decay after quantum-sensing-based cooling}
\begin{figure*}[ht]    \centering
    \includegraphics[width=1\textwidth]{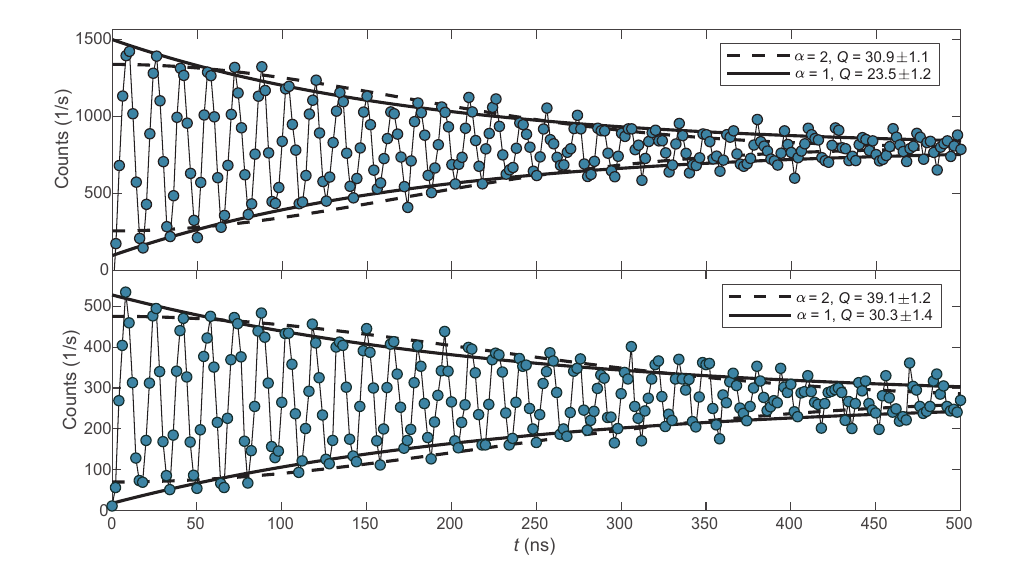}
    \caption{\textbf{Quality factor after quantum-sensing-based-cooling.} Comparison of the Rabi decay fitted to $\exp[-(t/T_2^\text{Rabi})^\alpha]$ for Gaussian (\mbox{$\alpha$=2}) and exponential (\mbox{$\alpha=1$}) decay. Rabi oscillations for \mbox{$\Omega=2\pi\times\SI{64.6}{\mega\Hz}$} without cooling (top), and after quantum-sensing-based cooling (bottom).}
    \label{suppfig:Rabi decay after cooling}
\end{figure*}

The decay envelope of the Rabi oscillations can take different forms \cite{stano_review_2022,yoneda_fast_2014}. We fit the data to $\exp[(-t/T_2^{\text{Rabi}})^{\alpha}]$ with $\alpha$ a free fit parameter. However, it is difficult to distinguish unambiguously between a Gaussian (\mbox{$\alpha =2$}) and an exponential (\mbox{$\alpha =1$}) decay: the residuals of the Gaussian and exponential fits are very similar. We decided to use the exponential fit to make a comparison to previous experiments \cite{bodey_optical_2019a,nakajima_coherence_2020}. Comparing Rabi oscillations with and without quantum-sensing-based cooling (Fig.~\ref{suppfig:Rabi decay after cooling}(a,b)) we find an increase in the quality factor from \mbox{$Q=23.5$} to $30.3$ consistent with a reduction of hyperfine-interaction-related decay. 

\section{Decoupling Sequences}
\begin{figure*}[ht]    \centering
    \includegraphics[width=1\textwidth]{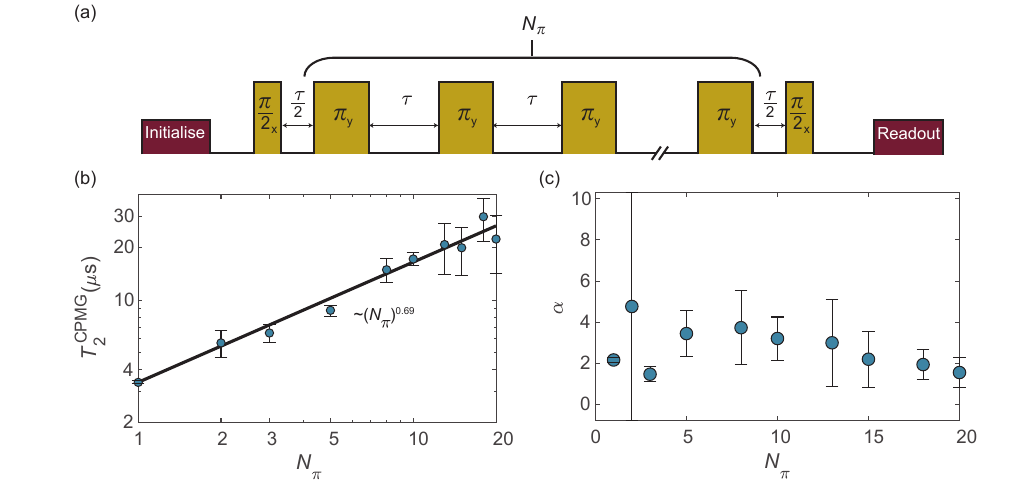}
    \caption{\textbf{CPMG decoupling pulse scaling.} (a) Pulse sequence for the CPMG experiment. (b) Electron spin decoupling using CPMG pulses. Solid line is a power-law fit with $T_2$ time scaling $T_2^{\text{CPMG}}\propto N_\pi^\gamma$ with \mbox{$\gamma=\SI{0.69\pm0.12}{}$}. (c) $\alpha$ extracted from a stretched exponential fit to the data shown in Fig.~\ref{suppfig:CPMG_data}. }
    \label{suppfig:CPMG}
\end{figure*}

\begin{figure*}[ht]    \centering
    \includegraphics[width=1\textwidth]{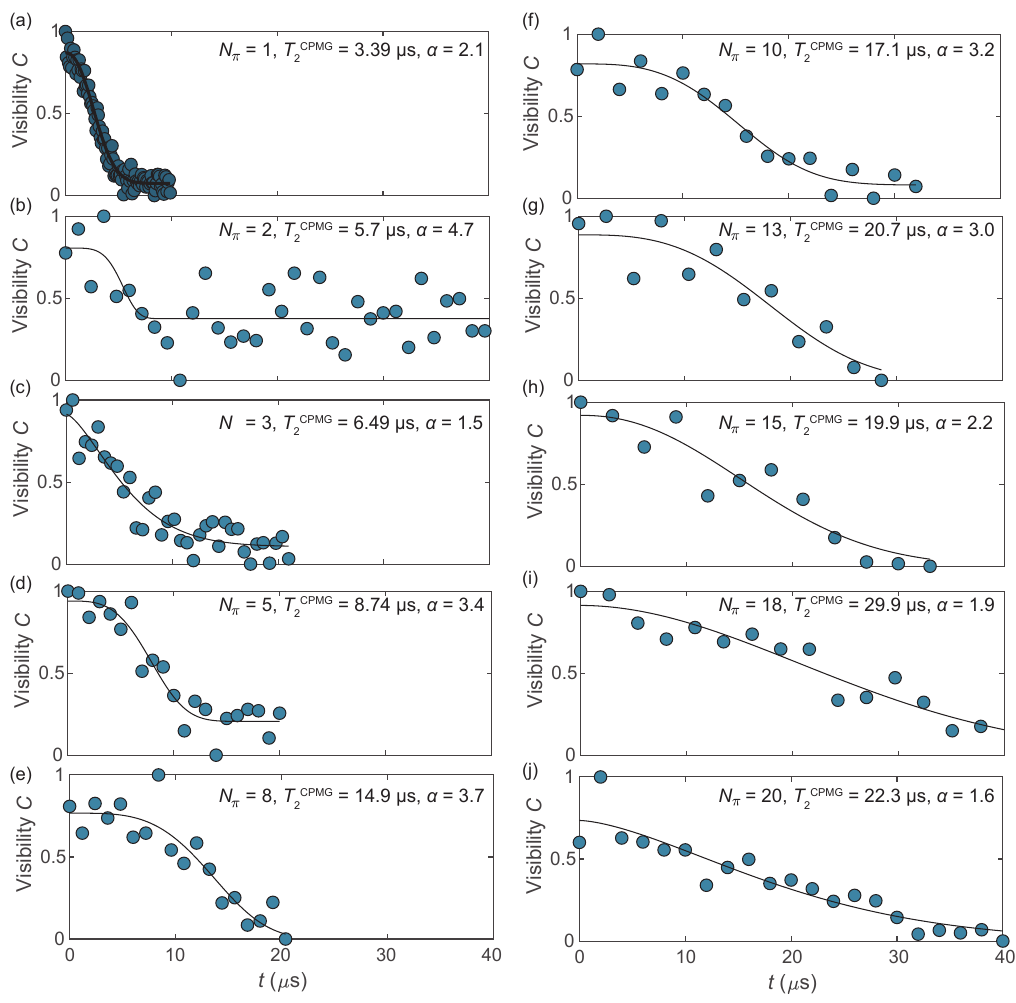}
    \caption{\textbf{CPMG decoupling data.} Electron spin decoupling using CPMG pulses. Solid lines are fits to stretched-exponentials $\exp[(-t/T_2^\text{CPMG})^\alpha]$.}
    \label{suppfig:CPMG_data}
\end{figure*}

The CPMG pulse sequence potentially allows the electron spin to be decoupled from nuclear noise, enhancing the spin $T_2$ time \cite{medford_scaling_2012}. The CPMG sequence is shown in Fig.~\ref{suppfig:CPMG}(a): after a $\frac{\pi}{2}_x$-pulse and $\tau/2$ delay, $N_\pi$ $\pi_y$-refocusing pulses separated by $\tau$ were applied followed by another $\tau/2$ delay and $\frac{\pi}{2}_x$-pulse. We fit the experimental data (see Fig.~\ref{suppfig:CPMG_data}) with stretched exponentials $\exp[(-t/T_2^{\text{CPMG}})^\alpha]$ and plot $T_2^{\text{CPMG}}$ against the number of applied refocusing pulses $N_\pi$, where \mbox{$t=2N_\pi\tau$}. Using a power-law fit, we find a $T_2$ time scaling of $N_\pi^\gamma$ with \mbox{$\gamma=\SI{0.69\pm0.12}{}$} (see Fig. \ref{suppfig:CPMG}(b,c)). The power-law fit is insensitive to the exponent $\alpha$ of the stretched exponential and gives insight into the noise sources in the environment \cite{medford_scaling_2012}. \mbox{$\gamma=\SI{0.69\pm0.12}{}$} is similar to the value obtained in Ref.~\cite{zaporski_ideal_2023a} (\mbox{$\gamma=0.75$}) for which up to 81 decoupling pulses were applied to droplet-etched GaAs QDs. Ref.~\cite{malinowski_spectrum_2017}
reports \mbox{$\gamma=0.8$} on gate-defined GaAs QDs for \mbox{$N_\pi<16$} and a crossover to \mbox{$\gamma=0.5$} for large $N_\pi$. On increasing $N_\pi$, no change of $\gamma$ was observed in the experiments presented here. 

The application of more refocusing pulses was not possible due to a reduction in readout contrast as more pulses were applied. This is due to imperfect pulse calibrations and the limit imposed by the electron-spin relaxation time $T_1$. These problems can be mitigated with more precise calibration and with a new sample with a larger tunnel barrier to the back contact, respectively.

\clearpage







\appendix*

\end{document}